\documentclass[prb,twocolumn,superscriptaddress,floatfix]{revtex4-1}
\usepackage[english]{babel}
\usepackage{graphicx}
\usepackage{amsmath}
\usepackage{amssymb}
\newcommand{\im}{\mathrm{i}}
\newcommand{\dif}{\mathrm{d}}
\def\hr{{\hat{\rho}(t)}}
\def\hf{{\hat{f}(t)}}
\newcommand{\e}{\mathrm{e}}
\newcommand{\mrd}{\mbox{\hspace*{0.6mm}}}
\begin{document}
\title{Non-equilibrium Optical Conductivity in Materials with Localized Electronic States}
\author{Veljko Jankovi\'c} \email{veljko.jankovic@ipb.ac.rs}
\affiliation{Scientific Computing Laboratory, Institute of Physics Belgrade, University of Belgrade, 
Pregrevica 118, 11080 Belgrade, Serbia}
\author{Nenad Vukmirovi\'c}\email{nenad.vukmirovic@ipb.ac.rs}
\affiliation{Scientific Computing Laboratory, Institute of Physics Belgrade, University of Belgrade, 
Pregrevica 118, 11080 Belgrade, Serbia}

\begin{abstract} 
A wide range of disordered materials contain electronic states that are spatially well localized.  In this work, we investigated the electrical response of such systems in non-equilibrium conditions to external electromagnetic field. {We obtained the expression for optical conductivity valid for any non-equilibrium state of electronic subsystem. In the case of incoherent non-equilibrium state, this expression contains only} the positions of localized electronic states, Fermi's golden rule transition probabilities between the states and the populations of electronic states. The same form of expression is valid both in the case of weak electron-phonon interaction and weak electron-impurity interaction that act as perturbations of electronic Hamiltonian. The derivation was performed by expanding the general expression for AC conductivity in powers of small 
electron-phonon interaction or electron-impurity interaction parameter. Applications of the expression to two model systems, a simple one-
dimensional Gaussian disorder model and the model of a realistic three-dimensional organic polymer material, were presented, as well.

\end{abstract} 


\maketitle{}

\section{Introduction}\label{Sec:Intro}
Electronic transport in semiconductors has been attracting significant research attention for more than half a century. Particular classes of semiconductors where interesting physical effects arise in electronic transport are semiconductors where a certain type of disorder is present in the system which leads to localization of electronic states. These include amorphous inorganic semiconductors (such as amorphous Si or Ge),\cite{Mott} inorganic crystals doped with randomly positioned impurities\cite{Shklovskii,Bottger} and organic semiconductors based on conjugated polymers or small molecules.\cite{Brutting,Sun,Pope,Schwoerer,Kampen,Klauk,Baranovskii} The latter class of materials triggered a particular interest in the last two decades due to their low production cost, which led to the development of a variety of organic electronic devices.\cite{nat401-685,nmat5-222,nat347-539,nmat7-376,apl62-585,apl90-142109,nmat5-197}

There is currently a solid understanding of equilibrium electronic transport in disordered systems with localized electronic states. DC transport in such systems can be modeled using an equivalent network of resistors that connect each two sites where electronic states are localized.\cite{pr120-745,prb4-2612} Electronic conductivity or mobility in the material can then be calculated by finding the equivalent resistance of the network or estimated using percolation theory. However, DC mobility which quantifies electronic transport properties over long length scales is in many cases not the most relevant quantity when the description of electronic transport processes is concerned. In particular, in organic solar cells based on a bulk heterojunction of two organic semiconductors, charge carriers travel over very short length (on the order of nanometers) and time (on the order of picoseconds) scales before they reach the interface of two semiconductors.\cite{sci258-1474,jpcc112-4350,afm22-1116} The high 
frequency (terahertz) AC mobility is a much better measure of charge transport over such short time 
scales.

The approaches for simulation of AC conductivity are usually based on Kubo's formula which expresses the AC conductivity in terms of the mean square displacement of a diffusing carrier.\cite{prb7-4491,prb83-081202,prb89-235201,irpc27-87,jppa215-123} Such approaches therefore assume that carriers are in equilibrium and that they are only slightly perturbed by external alternating electric field. However, in many realistic situations, the carriers are not in equilibrium; a typical example concerns the carriers created by external optical excitation across the band gap of a semiconductor. While general approaches for the treatment of non-equilibrium electronic transport, such as the density matrix formalism\cite{RevModPhys.74.895} or the non-equilibrium Green's function formalism,\cite{Keldysh,Kadanoff,Haug} do exist, it is in practice quite difficult to apply them to disordered materials, where one needs to consider large portions of material to obtain reliable information about its properties.

The main goal of this work was to derive a simple expression that relates the optical conductivity of a material with localized electronic states to its microscopic parameters. To accomplish this goal, we first derive in Sec.~\ref{Sec:geneoc} the relation between non-equilibrium optical conductivity and the corresponding current-current correlation function. Then, in Sec.~\ref{Sec:oceph} we derive an expression for the conductivity of the system of localized states with electron-phonon interaction that acts as perturbation {valid for arbitrary non-equilibrium state of the electronic subsystem. In the case of incoherent non-equilibrium state,} the obtained expression appears to have a rather simple form -- the only quantities that appear in it are the positions of localized states, their populations and the phonon-induced transition probabilities between the states. In Sec.~\ref{Sec:ocimp} we show that the same expression is obtained if additional static potential acts as a 
perturbation. In Sec.~\ref{Sec:num}, we 
present the results 
obtained from the application of the derived formula to a simple one-dimensional hopping model and to a realistic disordered conjugated polymer material. We discuss our results in light of the other results that exist in the literature in Sec.~\ref{Sec:disc}. 

\section{General expression for non-equilibrium optical conductivity}\label{Sec:geneoc}  

In this section, we consider an arbitrary quantum system described by the Hamiltonian $\hat H$ whose state is given by the statistical operator $\hat\rho(t)$. We will derive the time evolution of $\hat\rho(t)$ due to a weak external perturbation $\hat H'(t)$  which is turned on at $t=0$. Next, for the system that contains charged particles, we will find the current density caused by external electric field that acts as a perturbation. While the results of this section are mostly available in the literature, we repeat them here for completeness of the paper, as well as to introduce the notation and terminology for the remainder of the paper.

\subsection{Evolution of the density matrix}\label{Sec:th-edm}

The equation for the density matrix $\hat\rho(t)$ describing the state of the
system for $t>0$ is
\begin{equation}
 \im\hbar\frac{\dif\hat{\rho}(t)}{\dif t}=[\hat{H}+\hat{H}'(t),\hat \rho(t)]. \label{liuvil}
\end{equation}
We search for the solution of Eq.~(\ref{liuvil}) in the form $\hat\rho(t)=\hat{\rho}_\mathrm{free}(t)+\hat{f}(t)$, where 
\begin{equation}
\hat\rho_\mathrm{free}(t)=\e^{-\im \frac{\hat{H}}{\hbar}t}\hat{\rho}(0)\e^{\im\frac{\hat{H}}{\hbar}t}                                                    
\end{equation}                   
is the statistical operator of the system in the absence of external perturbation, $\hat\rho(0)$ is the statistical operator describing the state of the system just before the external perturbation is turned on, while $\hat f(t)$ is the contribution to the statistical operator due to linear response of the system. It satisfies the differential equation
\begin{equation}
\im\hbar\frac{\dif\hf}{\dif t}-[\hat{H},\hf]=[\hat{H}'(t),\hat{\rho}_\mathrm{free}(t)],
\end{equation}
with the initial condition $\hat f(0)=\hat 0$. After solving the last equation up to linear terms, we obtain\cite{Rammer}
\begin{equation}
\label{rhot}
\hat\rho(t)=\e^{-\im\frac{\hat{H}}{\hbar}t}\hat{\rho}(0)\e^{\im\frac{\hat{H}}{\hbar}t}+
\frac{1}{\im\hbar}
\e^{-\im\frac{\hat{H}}{\hbar}t}
\int_{0}^{t}\dif t'
[\hat{H}'_I(t'),\hat{\rho}(0)]
\e^{\im\frac{\hat{H}}{\hbar}t},
\end{equation}
where
$\hat{H}'_I(t)=\e^{\im\frac{\hat{H}}{\hbar}t}\hat{H}'(t)\e^{-\im\frac{\hat{H}}{\hbar}t}.$
Equations given in this section are strictly valid only for an isolated quantum system and do not include the relaxation of the system from some non-equilibrium to the equilibrium state. In realistic systems, interaction of the system with the environment leads to relaxation of the system to the equilibrium state. Therefore, the equations that we will derive are valid only if the characteristic time of  external perturbation is short compared to the relaxation time $\tau$. Since we shall study the response to the electric field oscillating with a frequency $\omega$, the aforementioned condition reads
\begin{equation}
\label{relax}
 \omega\tau\gg 1.
\end{equation}
In that case, the relaxation of the system towards equilibrium during one period of the perturbation is negligible and can be ignored in the considerations. 

\subsection{Non-equilibrium optical conductivity}\label{Sec:neoc}
Next, we assume that the system contains mobile charged particles and that external electric field acts as a perturbation. The system responds to external electric field by non$-$zero value of current density. The current density operator is given by~\cite{Mahan}
\begin{equation}
 \hat{j}_a(\mathbf{r})=\frac{1}{2m}\sum_{n}q\left(\hat{\mathbf{p}}_n\delta^{(3)}(\mathbf{r}-\hat{\mathbf{r}}_n)+
\delta^{(3)}(\mathbf{r}-\hat{\mathbf{r}}_n)\hat{\mathbf{p}}_n\right)_a,
\end{equation}
where $q$ and $m$ are the charge and the mass of a carrier, respectively, while $\hat{\mathbf{p}}_n$ and
$\hat{\mathbf{r}}_n$ are the momentum and the position operator for a single carrier, and $a$ denotes the component of the current density operator ($x$, $y$ or $z$). Using Eq.~\eqref{rhot},
we find that the current density at time $t$ is given as
\begin{eqnarray}\label{celastruja}
 \langle\hat{j}_a(\mathbf{r})\rangle_{t}&=&\mathrm{Tr}\left(\hr\hat{j}_a(\mathbf{r})\right)=\nonumber\\
&=&\mathrm{Tr}\left(\hat{\rho}(0)\hat{j}_a(t,\mathbf{r})\right)+ \\
&&+\frac{1}{\im\hbar}\int_0^t\dif t'\mrd\mathrm{Tr}\left(\hat{\rho}(0)[\hat{j}_a(t,\mathbf{r}),\hat{H}'_I(t')]\right), \nonumber
\end{eqnarray}
where $\hat{j}_a(t,\mathbf{r})=\e^{\im\frac{\hat{H}}{\hbar}t}\hat{j}_a(\mathbf{r})\e^{-\im\frac{\hat{H}}{\hbar}t}$.
The first term in Eq.~\eqref{celastruja} does not depend on the electric field, while we are interested in the response of the system to the applied electric field. Consequently, we shall further only consider the second term in Eq.~\eqref{celastruja} given as 
\begin{equation}
\label{deostruje}
 \mathcal{J}_a(t,\mathbf{r})=\frac{1}{\im\hbar}\int_0^t\dif t'\mrd\mathrm{Tr}
\left(\hat\rho(0)[\hat{j}_a(t,\mathbf{r}),\hat{H}'_I(t')]\right).
\end{equation}
We shall also assume that we are dealing with spatially homogeneous system at the macroscopic scale. The current density averaged over the volume of the system 
\begin{equation}
 \label{odgovor}
 \mathcal{J}_a(t)=\frac{1}{V}\int\dif^3\mathbf{r}\mrd\mathcal{J}_a(t,\mathbf{r})
\end{equation}
will be considered as the response to the applied field.
The Hamiltonian of interaction with electric field $\mathbf{E}(t)$ is given as~\cite{Jacoboni}
\begin{equation}
\label{interakcionih}
 \hat{H}'(t)=-\hat{\mathbf{\Pi}} \cdot \mathbf{E}(t),
\end{equation}
where $\hat{\mathbf{\Pi}}$ is the electric dipole moment operator defined as
\begin{equation}
 \label{dipolni}
 \hat{\mathbf{\Pi}}=q\sum_n\hat{\mathbf{r}}_n.
\end{equation}
Using Eqs.~\eqref{odgovor},~\eqref{deostruje} and~\eqref{interakcionih} the quantity $\mathcal{J}_a(t)$
can be expressed as
\begin{equation}
\label{eld1}
 \mathcal{J}_a(t)=\int_0^t\dif t'
\sigma_{ab}(t,t')E_b(t'),
\end{equation}
where the tensor
\begin{equation}
\label{linearniodzivvreme}
 \sigma_{ab}(t,t')=\frac{\im}{\hbar V}\mathrm{Tr}
\left(\hat{\rho}(0)\left[\hat J_a(t),\hat{\Pi}_b(t')\right]\right)
\end{equation}
describes the linear response to the applied electric field. In Eq.~\eqref{linearniodzivvreme}, the operator $\hat J_a(t)$ is defined as 
\begin{equation}
\label{uvodjenjebrzine}
 \hat J_a(t)=\int\dif^3\mathbf{r}\mrd\hat{j}_a(t,\mathbf{r})=\frac{q}{m}\sum_n \left(\mathbf{\hat{p}}_{n}\right)_a,
\end{equation}
The operators $\hat{\Pi}_a(t)$ i $\hat{J}_b(t)$ satisfy the equal time commutation relation
\begin{equation}
 \label{etcr}
 [\hat{\Pi}_a(t),\hat{J}_b(t)]=\im\hbar\frac{Nq^2}{m}\delta_{ab},
\end{equation}
where $N$ is the number of carriers, and the continuity equation 
\begin{equation}
 \label{jednkont}
 \hat{J}_a(t)=\frac{\dif}{\dif t}\hat{\Pi}_a(t).
\end{equation}
When the condition
\begin{equation}
 \label{uslov}
 [\hat \rho(0),\hat H]=\hat 0
\end{equation}
is satisfied, the tensor $\sigma_{ab}(t,t')$ defined in Eq.~\eqref{linearniodzivvreme} does not depend separately on $t$
and $t'$, but only on their difference $u=t-t'$. The optical conductivity tensor can then be defined as $\sigma_{ab}(\omega)=\int_0^{+\infty}\dif u\mrd\sigma_{ab}(u)\mrd\e^{\im\omega u}$ and reads
\begin{equation}
\label{sigmaab}
 \sigma_{ab}(\omega)=\frac{\im}{\hbar V}\int_0^{+\infty}\dif t\mrd\e^{\im\omega t}
\mrd\mathrm{Tr}\left(\hat{\rho}(0)[\hat{J}_a(t),\hat{\Pi}_b(0)]\right).
\end{equation}
Using Eqs.~\eqref{jednkont} and \eqref{etcr}, Eq.~\eqref{sigmaab} can be cast into a more familiar form ($n=N/V$ is the concentration of carriers)
\begin{equation}
\label{kubojj}
\begin{split}
 \sigma_{ab}(\omega)=\im\frac{nq^2}{m\omega}\delta_{ab}+
\frac{1}{\hbar\omega V}\int_0^{+\infty}\dif t\mrd\e^{\im\omega t}\times \\ \times
\mathrm{Tr}
\left(\hat{\rho}(0)[\hat{J}_a(t),\hat{J}_b(0)]\right).
\end{split}
\end{equation}
The equation for the optical conductivity~\eqref{kubojj} can be considered as a generalization to the non-equilibrium stationary case of well$-$known results\cite{allencontribution} which relate optical conductivity to the equilibrium current$-$current correlation function. 
Generalizations of this sort have already been proposed in the literature (in the context of the fluctuation$-$dissipation theorem, which relates the dissipative part of the optical conductivity to the (non$-$)equilibrium current fluctuations).\cite{arxiv0404270,prl93-250601}

In the case, when $[\hat \rho(0),\hat H]\neq\hat 0$, the tensor $\sigma_{ab}(t,t')$ defined in Eq.~\eqref{linearniodzivvreme} depends separately on $t$ and $t'$
\begin{equation}
\label{ponovljena}
\begin{split}
 \sigma_{ab}(t,t')=-\frac{1}{\im\hbar V}\mathrm{Tr}\left(
\e^{-\frac{\im}{\hbar}\hat H t}\hat\rho(0)\e^{\frac{\im}{\hbar}\hat H t} \right. \times \\ \times \left.
\left[\hat{J}_a(0),\hat\Pi_b\left(-(t-t')\right)\right]
\right).
\end{split}
\end{equation}
Using Eqs.~\eqref{jednkont} and \eqref{etcr}, one obtains the following expression:
\begin{equation}
\label{startnapozicija}
\begin{split}
 \sigma_{ab}(t,t')&=\frac{nq^2}{m}\mrd\delta_{ab}
-\frac{\im}{\hbar V}\int_0^{t-t'}\dif\tau \times \\
  \times & \mathrm{Tr}\left(
\e^{-\frac{\im}{\hbar}\hat H (t-\tau)}\hat\rho(0)\e^{\frac{\im}{\hbar}\hat H (t-\tau)}
\left[\hat{J}_a(\tau),\hat{J}_b(0)\right]
\right). 
\end{split}
\end{equation}

\section{Optical conductivity in the presence of electron-phonon interaction}\label{Sec:oceph}
In this section, we derive the expression for optical conductivity of a system with localized electronic states in the presence of weak electron-phonon interaction.
In Sec.~\ref{Sec:oceph1} we introduce the Hamiltonian of the system, derive the current operator
{and obtain the frequency$-$time representation of the conductivity tensor}.
In Sec.~\ref{Sec:oceph2}, we derive the expression {for conductivity valid for the arbitrary reduced density matrix of the electronic subsystem and} in the limit of low carrier concentration. {For incoherent density matrix of the electronic subsystem, this expression contains only the} 
 populations of single-particle electronic states, their spatial positions 
and Fermi's golden rule transition probabilities between these states.

\subsection{Model Hamiltonian and Preliminaries}\label{Sec:oceph1}

We consider the system of electrons and phonons described by the Hamiltonian
\begin{equation}
\label{modeleph}
 \hat H=\hat H_{0}+\hat H_{\mathrm{e-ph}}=\hat H_{\mathrm e}+\hat H_{\mathrm{ph}}+\hat H_{\mathrm{e-ph}},
\end{equation}
where
\begin{equation}
 \hat H_{0}=\hat H_{\mathrm e}+\hat H_{\mathrm{ph}}=\sum_{\alpha}\epsilon_\alpha\hat{c}_\alpha^\dagger\hat{c}_\alpha+
\sum_{k}\hbar\omega_{k}\mrd\hat{b}_{k}^\dagger\mrd\hat{b}_{k}
\end{equation}
is the Hamiltonian of noninteracting electrons and phonons, while 
\begin{equation}
 \hat H_{\mathrm{e-ph}}=
\sum_{k}\sum_{\alpha\alpha'}
\left(g_{\alpha\alpha',k}^{-}\mrd\hat{c}_\alpha^\dagger\hat{c}_{\alpha'}\hat{b}_{k}+
g_{\alpha\alpha',k}^{+}\mrd\hat{c}_{\alpha}^\dagger\hat{c}_{\alpha'}\hat{b}_{k}^\dagger\right)
\end{equation}
is the electron$-$phonon interaction Hamiltonian. In previous expressions, $\hat{b}_{k}^\dagger$ ($\hat{b}_{k}$) are the creation (annihilation) operators for the phonon mode $k$ that satisfy bosonic commutation relations, $\hat{c}_\alpha^\dagger$ ($\hat{c}_\alpha$) are creation (annihilation) operators for electronic single-particle state $\alpha$ that satisfy fermion anticommutation relations, $\hbar\omega_k$ is the energy of a mode $k$ phonon, while $\epsilon_\alpha$ is the energy of electronic state $\alpha$. Matrix elements of electron$-$phonon interaction satisfy the relation
\begin{equation}
 g_{\alpha\alpha',k}^{\pm}=g_{\alpha'\alpha,k}^{\mp*}
\end{equation}
and their particular form depends on details of electron$-$phonon interaction mechanism.

We will assume that phonon subsystem is in thermal equilibrium and therefore we will adopt the
following factorization of the initial density matrix $\hat\rho(0)$
\begin{equation}
\label{statfakt}
 \hat\rho(0)=\hat\rho_\mathrm{e}\mrd\hat\rho_\mathrm{ph,eq}.
\end{equation}
The operator $\hat\rho_\mathrm{ph,eq}$ describes the phonon subsystem in equilibrium at the temperature $T_\mathrm{ph}=\frac{1}{k_B\beta_\mathrm{ph}}$ and it is given as
\begin{equation}
 \hat\rho_\mathrm{ph,eq}=
\frac{\e^{-\beta_{\mathrm{ph}}\hat H_\mathrm{ph}}}{\mathrm{Tr}_\mathrm{ph}\mrd\e^{-\beta_{\mathrm{ph}}\hat H_\mathrm{ph}}},
\end{equation}
whereas $\hat\rho_\mathrm{e}$ is the reduced density matrix of the electronic subsystem. The state of the system described by Eq.~(\ref{statfakt}) assumes that electrons are out of equilibrium, while the phonons are in equilibrium. Such states can arise naturally in several relevant physical scenarios. Typical example of such a scenario is a semiconductor structure excited with photons whose energy is larger than the band gap of the structure. 
Most of the energy of incident photons is then transferred to electronic degrees of freedom and therefore it is quite reasonable to assume that electrons are out of equilibrium, while the phonons are in equilibrium.

The electric dipole moment operator $\hat{\Pi}_a$ introduced in Eq.~\eqref{dipolni} can be expressed in the second
quantization representation as
\begin{equation}
 \hat\Pi_a=q\sum_{\alpha\beta}x_{a;\alpha\beta}\mrd\hat c_\alpha^\dagger\hat c_{\beta},
\end{equation}
where $ x_{a;\alpha\beta}\equiv\langle\alpha|\hat x_a|\beta\rangle$ are the matrix elements of the single electron position operator.
Using Eq.~\eqref{jednkont}, we find that the operator $\hat J_a$ reads
$\hat{J}_a=\hat{J}_a^{(1)}+\hat{J}_a^{(2)}$, where
\begin{equation}
 \hat{J}_a^{(1)}=\frac{\im q}{\hbar}\sum_{\alpha\beta}(\epsilon_\alpha-\epsilon_\beta)x_{a;\alpha\beta}\mrd
\hat c_\alpha^\dagger\hat c_\beta
\end{equation}
describes the contribution to the operator $\hat J_a$ due to direct interaction of electrons with electric field, while
\begin{equation}
\label{Ja2}
 \hat J_a^{(2)}=\frac{\im q}{\hbar}\sum_{k}\sum_{\alpha\beta}\left(
F_{a;\alpha\beta,k}^-\mrd\hat c_\alpha^\dagger\hat c_\beta\hat b_k+
F_{a;\alpha\beta,k}^+\mrd\hat c_\alpha^\dagger\hat c_\beta\hat b_k^\dagger\right)
\end{equation}
describes the contribution arising from electron$-$phonon interaction. The coefficients $F_{a;\alpha\beta,k}^\pm$
are given by
\begin{equation}
\label{deff}
 F_{a;\alpha\beta,k}^\pm=
\sum_{\alpha'}\left(
g_{\alpha\alpha',k}^{\pm}\mrd x_{a;\alpha'\beta}-
x_{a;\alpha\alpha'}g_{\alpha'\beta,k}^{\pm}\right)
\end{equation}
and satisfy the relation
\begin{equation}
\label{osobinef}
 F_{a;\alpha\beta,k}^\pm=-F_{a;\beta\alpha,k}^{\mp*}.
\end{equation}
In this work, we are mainly interested in the case of localized electronic states when the matrix elements of the position operator between different states are negligible. This condition can be mathematically expressed as \begin{equation}
\label{deflok}
 x_{a;\alpha\beta}=\delta_{\alpha\beta}\mrd x_{a;\alpha}.
\end{equation}
Therefore, in the case of localized electronic states, $\hat{J}_a^{(1)}=0$ and consequently $\hat{J}_a=\hat{J}_a^{(2)}$.

Next, we treat electron-phonon interaction as a perturbation and perform the expansion of Eq.~\eqref{startnapozicija} with respect to small interaction constants $g_{\alpha\beta,k}^\pm$. The evolution operator that appears in Eq.~\eqref{startnapozicija} can be expanded in Dyson series as
\begin{equation}
\label{dyson}
 \e^{-\frac{\im}{\hbar}\hat H t}=\e^{-\frac{\im}{\hbar}\hat H_{0}t}+
\frac{1}{\im\hbar}\int_0^{t}\dif t'\mrd\e^{-\frac{\im}{\hbar}\hat H_{0}(t-t')}
\hat H_{\mathrm{e-ph}}\e^{-\frac{\im}{\hbar}\hat H_{0}t'}+\dots
\end{equation}
Consequently, the expansion of the time-dependent operator $\hat J_a(\tau)$ from Eq.~\eqref{startnapozicija} reads
\begin{eqnarray}
\label{strujarazvoj}
 &&\hat J_a(\tau)
=\e^{\frac{\im}{\hbar}\hat H_{0}\tau}\hat J_a\e^{-\frac{\im}{\hbar}\hat H_{0}\tau}+\\
&&+\left[\e^{\frac{\im}{\hbar}\hat H_{0}\tau}\hat J_a\e^{-\frac{\im}{\hbar}\hat H_{0}\tau},\int_0^\tau
\frac{\dif t'}{\im\hbar}\e^{\frac{\im}{\hbar}\hat H_{0}t'}\hat H_{\mathrm{e-ph}}\e^{-\frac{\im}{\hbar}\hat H_{0}t'}\right]
+\dots    \nonumber
\end{eqnarray} 
Furthermore, the expansion of the first term under trace in Eq.~\eqref{startnapozicija} gives
\begin{equation}
\label{bach}
\begin{split}
 \e^{-\frac{\im}{\hbar}\hat H (t-\tau)}\hat\rho(0)\e^{\frac{\im}{\hbar}\hat H (t-\tau)}=
\hat\rho(0)+\\+\sum_{n=1}^{+\infty}\frac{1}{n!}\left(-\frac{\im(t-\tau)}{\hbar}\right)^n
[\hat H,\dots,[\hat H,\hat\rho(0)\underbrace{]\dots]}_n
\end{split}
\end{equation}
Our aim is to obtain the first non-zero term in the expansion of Eq.~\eqref{startnapozicija} in the case of localized electronic states.
It is therefore sufficient to take only the first term in the expansion given by Eq.~\eqref{strujarazvoj} and to isolate the contribution
from the expansion given in Eq.~\eqref{bach} which does not contain electron-phonon coupling constants.
One can show by direct inspection, using the factorization of the initial density matrix given by Eq.~\eqref{statfakt},
that every summand under the sum on the right hand side of Eq.~\eqref{bach}
has only one term which does not contain electron-phonon coupling constants and which is of the type
$\displaystyle{
\frac{1}{n!}\left(-\frac{\im(t-\tau)}{\hbar}\right)^n
[\hat H_\mathrm{e},\dots,[\hat H_\mathrm{e},\hat\rho_\mathrm{e}\underbrace{]\dots]}_n\hat\rho_\mathrm{ph,eq}.
}$
All these contributions can be resummed so that we finally obtain the zeroth-order term in the expansion given by Eq.~\eqref{bach}
\begin{equation}
\label{bachnovi}
\begin{split}
 &\e^{-\frac{\im}{\hbar}\hat H (t-\tau)}\hat\rho(0)\e^{\frac{\im}{\hbar}\hat H (t-\tau)}=\\
&=\e^{-\frac{\im}{\hbar}\hat H_\mathrm{e} (t-\tau)}\hat\rho(0)\mrd\e^{\frac{\im}{\hbar}\hat H_\mathrm{e} (t-\tau)}
+\dots
\end{split}
\end{equation}
The first nontrivial term in the expansion of Eq.~\eqref{startnapozicija} in the case of localized electronic states is thus given by
\begin{equation}
\label{startnapozicijanova}
\begin{split}
 \sigma_{ab}(t,t')&=\frac{nq^2}{m}\mrd\delta_{ab}
-\frac{\im}{\hbar V}\int_0^{t-t'}\dif\tau \times \\
  \times & \mathrm{Tr}\left(
\hat\rho(0)\mrd
\e^{\frac{\im}{\hbar}\hat H_\mathrm{e} (t-\tau)}
\left[\hat{J}^{(2),0}_a(\tau),\hat{J}^{(2)}_b(0)\right]
\e^{-\frac{\im}{\hbar}\hat H_\mathrm{e} (t-\tau)}
\right)
\end{split}
\end{equation}
where $\hat{J}^{(2),0}_a(\tau)=\e^{\frac{\im}{\hbar}\hat H_0 \tau} \hat{J}^{(2)}_a 
\e^{-\frac{\im}{\hbar}\hat H_0 \tau}$. Next, we consider the frequency-time representation of the conductivity tensor which can be defined as
\begin{equation}
\label{Eq:mixed-rep}
 \sigma_{ab}(t,\omega)=\int_0^{+\infty}\dif u\mrd\sigma_{ab}(t,t-u)\mrd\e^{\im\omega u},
\end{equation}
When $\sigma_{ab}(t,t')$ depends only on the difference $t-t'$, Eq.~\eqref{Eq:mixed-rep}
defines the conventional optical conductivity tensor $\sigma_{ab}(\omega)$. The frequency-time representation of the conductivity tensor given in
Eq.~\eqref{startnapozicijanova} is
\begin{equation}
 \label{max2}
 \begin{split}
 \sigma_{ab}(t,\omega)=\frac{\im nq^2}{m\omega}\delta_{ab}+
\frac{1}{\hbar\omega V}\int_{0}^{+\infty}\dif u\mrd\e^{\im\omega u}
\mrd \times\\ \times \mathrm{Tr}\left(
\hat\rho(0)\mrd
\e^{\frac{\im}{\hbar}\hat H_\mathrm{e} (t-u)}
\left[\hat{J}^{(2),0}_a(u),\hat{J}^{(2)}_b\right]
\e^{-\frac{\im}{\hbar}\hat H_\mathrm{e} (t-u)}
\right).
\end{split}
\end{equation}
In the case when $\sigma_{ab}(t,\omega)$ varies slowly with $t$ on the $1/\omega$ timescale, it can be interpreted as the conventional optical conductivity tensor at time $t$.

We also note that when the condition of localized electronic states [Eq.~\eqref{deflok}] is not satisfied, the dominant term in the $\hat J_a$ operator is the $\hat{J}_a^{(1)}$ term. The leading terms in expansions \eqref{strujarazvoj} and \eqref{bach}
are then given by first terms in Eqs.~\eqref{strujarazvoj} and~\eqref{bachnovi}, where $\hat J_a$ is replaced by $\hat{J}_a^{(1)}$ in Eq.~\eqref{strujarazvoj}. These terms are independent of electron-phonon coupling constants and lead to the following expression
for the frequency-time representation of the conductivity tensor
\begin{equation}
 \label{max0}
 \begin{split}
 \sigma_{ab}(t,\omega)=\frac{\im nq^2}{m\omega}\delta_{ab}+
\frac{1}{\hbar\omega V}\int_{0}^{+\infty}\dif u\mrd\e^{\im\omega u}
\mrd \times\\ \times \mathrm{Tr}\left(
\hat\rho(0)\mrd
\e^{\frac{\im}{\hbar}\hat H_\mathrm{e} (t-u)}
\left[\hat{J}^{(1),0}_a(u),\hat{J}^{(1)}_b\right]
\e^{-\frac{\im}{\hbar}\hat H_\mathrm{e} (t-u)}
\right).
\end{split}
\end{equation}

The physical origin of this term is direct absorption of electromagnetic radiation by the electronic subsystem. However, since this term vanishes for a system with localized electronic states, which is of main interest in this work, this term will not be considered in the remainder of the paper. The focus will be on the term from Eq.~\eqref{max2} which arises due to phonon-assisted transitions between states, as will become evident in Sec~\ref{Sec:oceph2}.

\subsection{Frequency dependence of mobility in low carrier density limit}\label{Sec:oceph2}

We will now start from Eq.~\eqref{max2} for the frequency$-$time representation of conductivity to derive the expression for the optical conductivity that explicitly contains the populations of electronic states
{(diagonal elements of $\hat\rho_\mathrm{e}$) and coherences (off-diagonal elements of $\hat\rho_\mathrm{e}$)}
By replacing Eq.~\eqref{Ja2} into the expression for mean value 
$\displaystyle{\mathrm{Tr}\left(
\hat\rho(0)\mrd
\e^{\frac{\im}{\hbar}\hat H_\mathrm{e} (t-u)}
\left[\hat{J}^{(2),0}_a(u),\hat{J}^{(2)}_b\right]
\e^{-\frac{\im}{\hbar}\hat H_\mathrm{e} (t-u)}
\right)}$
and tracing out the phonon degrees of freedom one obtains
\begin{widetext}
\begin{equation}
\label{j22}
\begin{split}
&\mathrm{Tr}\left(
\hat\rho(0)\mrd
\e^{\frac{\im}{\hbar}\hat H_\mathrm{e} (t-u)}
\left[\hat{J}^{(2),0}_a(u),\hat{J}^{(2)}_b\right]
\e^{-\frac{\im}{\hbar}\hat H_\mathrm{e} (t-u)}
\right)\left(\frac{\im q}{\hbar}\right)^{-2}\\
=&\sum_{k}\sum_{\alpha\beta\gamma\delta}\left(
F_{a;\alpha\beta,k}^{-}F_{b;\gamma\delta,k}^{+}\mrd
\e^{-\frac{\im}{\hbar}(\epsilon_\gamma-\epsilon_\delta+\hbar\omega_k)u}-
F_{a;\alpha\beta,k}^{+}F_{b;\gamma\delta,k}^{-}\mrd
\e^{-\frac{\im}{\hbar}(\epsilon_\gamma-\epsilon_\delta-\hbar\omega_k)u}\right)
\e^{\frac{\im}{\hbar}(\epsilon_\alpha-\epsilon_\beta+\epsilon_\gamma-\epsilon_\delta)t}\mrd
\langle\hat c_{\alpha}^{\dagger}\hat c_{\beta}\hat c_{\gamma}^{\dagger}\hat c_{\delta}\rangle_{\mathrm e}\\
+&\sum_{k}\sum_{\alpha\beta\gamma}
\left(
F_{a;\alpha\gamma,k}^{-}F_{b;\gamma\beta,k}^{+}\mrd
\e^{-\frac{\im}{\hbar}(\epsilon_\gamma-\epsilon_\beta+\hbar\omega_k)u}-
F_{b;\alpha\gamma,k}^{+}F_{a;\gamma\beta,k}^{-}\mrd
\e^{-\frac{\im}{\hbar}(\epsilon_\alpha-\epsilon_\gamma+\hbar\omega_k)u}
\right)
\mrd N_{k}\mrd
\e^{\frac{\im}{\hbar}(\epsilon_\alpha-\epsilon_\beta)t}\mrd\langle\hat c_{\alpha}^{\dagger}\hat c_{\beta}\rangle_{\mathrm e}\\
+&\sum_{k}\sum_{\alpha\beta\gamma}
\left(
F_{a;\alpha\gamma,k}^{+}F_{b;\gamma\beta,k}^{-}\mrd
\e^{-\frac{\im}{\hbar}(\epsilon_\gamma-\epsilon_\beta-\hbar\omega_k)u}-
F_{b;\alpha\gamma,k}^{-}F_{a;\gamma\beta,k}^{+}\mrd
\e^{-\frac{\im}{\hbar}(\epsilon_\alpha-\epsilon_\gamma-\hbar\omega_k)u}
\right)
(1+N_{k})\mrd
\e^{\frac{\im}{\hbar}(\epsilon_\alpha-\epsilon_\beta)t}
\mrd\langle\hat c_{\alpha}^{\dagger}\hat c_{\beta}\rangle_{\mathrm e}.
\end{split}
\end{equation}
\end{widetext}
Here, $N_k$ is the number of phonons in mode $k$ given by the Bose-Einstein distribution, $\langle\dots\rangle_{\mathrm{e}}$ denotes averaging with respect to $\hat\rho_\mathrm{e}$ and coefficients $F_{\alpha\beta,k}^\pm$
are given as (by the virtue of the definition of the localized electronic states from Eq.~\eqref{deflok})
\begin{equation}
F_{a;\alpha\beta,k}^\pm=g_{\alpha\beta,k}^\pm(x_{a;\beta}-x_{a;\alpha}),
\end{equation}
% and satisfy
% \begin{equation}
% \label{osobine1d}
%  F_{\alpha\beta,k}^\pm=-F_{\beta\alpha,k}^{\mp*},
% \end{equation}
see Eqs.~\eqref{deff} and~\eqref{osobinef}.

In the limit of low carrier densities, only single-particle electronic excitations are relevant. One can therefore restrict the Hilbert space of the system to the space given as a product of single particle electronic space and the phonon space. In this restricted space, the operators $c_\alpha^\dagger\hat c_\beta$ and $c_\alpha^\dagger\hat c_\beta\hat c_\gamma^\dagger\hat c_\delta$ reduce respectively to $|\alpha\rangle\langle\beta|$ and $\delta_{\beta\gamma}|\alpha\rangle\langle\delta|$, while the Hamiltonian in this restricted space reads
\begin{equation}
\begin{split}
  \hat H&=\sum_{\alpha}\epsilon_\alpha|\alpha\rangle\langle\alpha|+
\sum_{k}\hbar\omega_{k}\mrd\hat{b}_{k}^\dagger\mrd\hat{b}_{k}+ \\
&+\sum_{k}\sum_{\alpha\alpha'}
\left(g_{\alpha\alpha',k}^{-}\mrd|\alpha\rangle\langle\alpha'|\hat{b}_{k}+
g_{\alpha\alpha',k}^{+}\mrd|\alpha\rangle\langle\alpha'|\hat{b}_{k}^\dagger\right).
\end{split}
\end{equation}
The average values of the expressions appearing in Eq.~\eqref{j22} are then given as
\begin{equation}
\label{2u1}
 \langle\hat c_\alpha^\dagger\hat c_\beta\rangle_\mathrm e=
\mathrm{Tr}_\mathrm e\left(\hat\rho_\mathrm e\hat c_\alpha^\dagger\hat c_\beta\right)
=\langle\beta|\hat\rho_\mathrm e|\alpha\rangle,
\end{equation}
and
\begin{equation}
\label{4u1}
 \langle\hat c_\alpha^\dagger\hat c_\beta\hat c_\gamma^\dagger\hat c_\delta\rangle_\mathrm e=
\mathrm{Tr}_\mathrm e\left(\hat\rho_\mathrm e\hat c_\alpha^\dagger\hat c_\beta\hat c_\gamma^\dagger\hat c_\delta\right)
=\delta_{\beta\gamma}
\langle\delta|\hat\rho_\mathrm e|\alpha\rangle.
\end{equation}

Combining Eqs.~\eqref{2u1},~\eqref{4u1},~\eqref{j22} and~\eqref{max2}, the following equation for the frequency-time representation of
the conductivity tensor is obtained
\begin{equation}
 \label{Eq:frequency-time}
 \sigma_{ab}(t,\omega)=\im\mrd\frac{nq^2}{m\omega}\mrd\delta_{ab}-f_{ab}(t,\omega)-f_{ab}(t,-\omega)^*
\end{equation}
where $f_{ab}(t,\omega)$ is defined as
\begin{equation}
\label{Eq:definition_f}
\begin{split}
 &f_{ab}(t,\omega)=\frac{q^2}{\hbar^2\omega V}\sum_{k}\sum_{\alpha\beta\gamma}\e^{\frac{\im}{\hbar}(\epsilon_\alpha-\epsilon_\beta)t}
\langle\beta|\hat\rho_\mathrm{e}|\alpha\rangle \times \\
&\times
\left(
F_{a;\alpha\gamma,k}^{-}F_{b;\gamma\beta,k}^{+}\mathcal{D}(\epsilon_\beta-\epsilon_\gamma-\hbar\omega_k+\hbar\omega)(1+N_k)+  \right. \\ & \left.+
F_{a;\alpha\gamma,k}^{+}F_{b;\gamma\beta,k}^{-}\mathcal{D}(\epsilon_\beta-\epsilon_\gamma+\hbar\omega_k+\hbar\omega) N_k\right).
\end{split}
\end{equation}
Function $\mathcal{D}(\epsilon)$ is given as
\begin{equation}
 \mathcal{D}(\epsilon)=\pi\delta(\epsilon)+\im\mrd\mathcal{P}(1/\epsilon).
\end{equation}
In the expressions \eqref{Eq:frequency-time} and \eqref{Eq:definition_f}, there are two clear signatures of non-equilibrium: the explicit time dependence and the presence of off-diagonal elements of $\hat\rho_\mathrm{e}$ (coherences). Both of these effects would be absent for the system in equilibrium.

Next, we consider the case when the reduced density matrix of the electronic subsystem is an analytic function of
the electronic Hamiltonian $\hat H_\mathrm{e}$, when we have
\begin{equation}
\label{analiticka}
 \langle\beta|\hat\rho_\mathrm e|\alpha\rangle=\delta_{\alpha\beta}\mrd r_\alpha,
\end{equation}
where
\begin{equation}
\label{ralpha}
 r_\alpha=\langle\alpha|\hat\rho_\mathrm e|\alpha\rangle
\end{equation}
is the average occupation of electronic state $\alpha$.
Then in Eq.~\eqref{Eq:definition_f} we remain only with the average occupations of
individual electronic states and since the quantity $f_{ab}(t,\omega)$ does not depend explicitly on $t$,
$\sigma_{ab}(t,\omega)$ also does not depend on $t$ and represents the frequency-dependent conductivity tensor.
Starting from Eqs.~\eqref{Eq:frequency-time} and~\eqref{Eq:definition_f}
one can show that under the aforementioned condition the following relation for the real part of the optical conductivity holds.

\begin{equation}
\label{opstislucajfzpneravnotezna}
\begin{split}
 \mathrm{Re}\mrd\sigma_{ab}&(\omega)=\frac{q^2}{2\hbar\omega V}\sum_{\alpha\beta}
(x_{a;\beta}-x_{a;\alpha})(x_{b;\beta}-x_{b;\alpha})
r_\beta \times \\ \times
&\left[w_{\beta\alpha,\mathrm{ph}}(\epsilon_\beta-\epsilon_\alpha+\hbar\omega)- 
      w_{\beta\alpha,\mathrm{ph}}(\epsilon_\beta-\epsilon_\alpha-\hbar\omega)\right].
\end{split}      
\end{equation}
where the terms $w_{\beta\alpha,\mathrm{ph}}$ are of the form
\begin{equation}
\label{femijevozp}
\begin{split}
 w_{\beta\alpha,\mathrm{ph}}(\epsilon_\beta-\epsilon_\alpha)=
\frac{2\pi}{\hbar}\sum_{k}   \left[
 |g_{\alpha\beta,k}^-|^2\delta(\epsilon_\beta-\epsilon_\alpha+\hbar\omega_{k})N_{k}+ \right. \\ \left. 
+|g_{\alpha\beta,k}^+|^2\delta(\epsilon_\beta-\epsilon_\alpha-\hbar\omega_{k})(1+N_{k})
\right].
\end{split}
\end{equation}
These are identical to the rates that would be obtained by applying Fermi's golden rule to calculate the transition probability from the state $\beta$ to the state $\alpha$ due to electron-phonon interaction. Eq.~\eqref{opstislucajfzpneravnotezna} gives a rather simple expression for the dissipative part of the optical conductivity as it involves the positions of electronic states, their occupations and the Fermi's golden rule transition probabilities. Eqs.~\eqref{opstislucajfzpneravnotezna} and \eqref{femijevozp} also offer an intuitive interpretation of elementary processes giving contribution to the dissipative part of the optical conductivity in the lowest non$-$trivial order of the perturbation expansion. These processes are one$-$particle transitions $\beta\to\alpha$ induced by emission (absorption) of one phonon accompanied by emission (absorption) of
the quantum of the external electromagnetic field $\hbar\omega$.
{One should note that within our lowest-order perturbative approach, one does not take into account multphonon transitions which may be important in some systems.}

From the definition of mobility, one then also obtains for the real part of AC mobility
\begin{equation}
\label{Eq:re-mu-eph}
\begin{split}
 \mathrm{Re}\mrd\mu_{ab}&(\omega)=\frac{q}{2\hbar\omega}\sum_{\alpha\beta}
(x_{a;\beta}-x_{a;\alpha})(x_{b;\beta}-x_{b;\alpha})
\frac{r_\beta}{\sum_\gamma r_\gamma} \times \\ \times
&\left[w_{\beta\alpha,\mathrm{ph}}(\epsilon_\beta-\epsilon_\alpha+\hbar\omega)- 
      w_{\beta\alpha,\mathrm{ph}}(\epsilon_\beta-\epsilon_\alpha-\hbar\omega)\right].
\end{split}      
\end{equation}
Eq. \eqref{Eq:re-mu-eph} was derived under the assumption that hopping rates have the mathematical form given by Eq.~\eqref{femijevozp}. In Sec.~\ref{Sec:num}, the hopping rates given by Eqs. \eqref{eq:yuk-aa} and \eqref{eq:ma_rates} will be used. Eq. \eqref{eq:yuk-aa} can be derived from Eq.~\eqref{femijevozp} under the assumption that electron-phonon coupling elements are proportional to wave function moduli overlap (see Ref. \onlinecite{apl97-043305}). Eq.~\eqref{eq:ma_rates} can then be obtained from Eq. \eqref{eq:yuk-aa} if one assumes that wave function overlaps decay exponentially with distance between states and that phonon density of states (DOS) is such that energy dependence in Eq. \eqref{eq:yuk-aa} disappears. Therefore both Eqs. \eqref{eq:yuk-aa} and \eqref{eq:ma_rates} are compatible with the mathematical structure of Eq.~\eqref{femijevozp} and it is appropriate to use them in Eq. \eqref{Eq:re-mu-eph}.

\section{Optical conductivity in the presence of impurity scattering}\label{Sec:ocimp}

In this section, we will show that {similar} expressions for optical conductivity are obtained if electrons interact with an additional static potential, rather than with phonons. A typical cause of such potential could be the impurities that are present in the material.

Therefore, we consider the Hamiltonian 
\begin{equation}
\label{impurityhamiltonian}
 \hat H=\hat H_{0}+\hat U=\sum_{\alpha}\epsilon_\alpha\hat c_\alpha^\dagger\hat c_\alpha+
\sum_{\alpha\beta}A_{\alpha\beta}\mrd\hat c_\alpha^\dagger\hat c_\beta,
\end{equation}
where $\hat H_{0}$ is non-interacting part of the Hamiltonian, while $\hat U$ describes the interaction of electrons with static potential.
The operator $\hat J_a$ can be computed using Eq.~\eqref{jednkont} and reads
\begin{equation}
\label{jaimp}
\begin{split}
 \hat J_a&=\hat J_a^{\mathrm{(dir)}}+\hat J_a^{\mathrm{(imp)}}=\\
&=\frac{\im q}{\hbar}\sum_{\alpha\alpha'}x_{a;\alpha\alpha'}(\epsilon_\alpha-\epsilon_{\alpha'})\hat c_\alpha^\dagger\hat c_{\alpha'}+
\frac{\im q}{\hbar}\sum_{\alpha\beta}\mathcal{A}_{a;\alpha\beta}\mrd\hat c_\alpha^\dagger\hat c_\beta.
\end{split}
\end{equation}
The $\hat J_a^{(\mathrm{dir})}$ operator is analogous to the operator $\hat J_a^{(1)}$ in the case of system with electron$-$phonon interaction and describes direct interaction of electrons with perturbing electric field. On the other hand, the $\hat J_a^{\mathrm{(imp)}}$ operator describes the contribution to $\hat J_a$ due to the interaction with the static potential (or, in particular, with impurities). The coefficients $\mathcal{A}_{a;\alpha\beta}$ that appear in Eq.~\eqref{jaimp} are given as
\begin{equation}
 \mathcal{A}_{a;\alpha\beta}=
\sum_{\alpha'}\left(A_{\alpha\alpha'}x_{a;\alpha'\beta}-
x_{a;\alpha\alpha'}A_{\alpha'\beta}\right)
\end{equation}
and satisfy (compare to Eq.~\eqref{osobinef})
\begin{equation}
 \mathcal{A}_{a;\beta\alpha}=-\mathcal{A}_{a;\alpha\beta}^*.
\end{equation}

We will treat the interaction with the static potential as a perturbation and we will derive the formula for optical conductivity in the lowest order of the perturbation expansion with respect to small coefficients $A_{\alpha\beta}$. We will assume that electronic states are localized, see Eq.~\eqref{deflok}. This way, the expression
for the operator $\hat J_a$ simplifies to
\begin{equation}
\label{jimp}
 \hat J_a=\hat J_a^{(\mathrm{imp})}=\frac{\im q}{\hbar}\sum_{\alpha\beta}A_{\alpha\beta}(x_{a;\beta}-x_{a;\alpha})\hat c_\alpha^\dagger\hat c_\beta.
\end{equation}

The starting
point for the perturbation expansion is again Eq.~\eqref{startnapozicija}.
Following a discussion, similar to that conducted in Sec.~\ref{Sec:oceph}, we obtain that the
first non$-$zero term in the expansion of Eq.~\eqref{startnapozicija} in the case of localized electronic states is
quadratic in quantities $A_{\alpha\beta}$ and that the corresponding expression for the time$-$frequency representation
of the conductivity tensor (Eq.~\eqref{Eq:mixed-rep}) reads
\begin{equation}
\label{max2imp}
\begin{split}
 \sigma_{ab}(t,\omega)=\frac{\im nq^2}{m\omega}\delta_{ab}+
\frac{1}{\hbar\omega V}\int_{0}^{+\infty}\dif t\mrd\e^{\im\omega t} \times \\ \times
\mrd\mathrm{Tr}\left(
\hat\rho(0)\mrd
\e^{\frac{\im}{\hbar}\hat H_0(t-u)}\left[\hat{J}^{(\mathrm{imp}),0}_a(u),\hat{J}^{(\mathrm{imp})}_b\right]
\e^{-\frac{\im}{\hbar}\hat H_0(t-u)}
\right).
\end{split}
\end{equation}
The notation $\hat{J}^{(\mathrm{imp}),0}_a(t)$ again suggests that the time dependence is governed by the non$-$interacting Hamiltonian.

{In the low density limit, the projection of the Hamiltonian onto the single-particle subspace reads
\begin{equation}
 \hat H_0=\sum_{\alpha}\epsilon_\alpha|\alpha\rangle\langle\alpha|+\sum_{\alpha\beta}A_{\alpha\beta}|\alpha\rangle\langle\beta|,
\end{equation}
with the average values $\langle\hat c_\alpha^\dagger\hat c_\beta\rangle=\langle\beta|\hat\rho(0)|\alpha\rangle$.
Deriving Eq.~\eqref{max2imp} we obtain the expression for the frequency$-$time representation of the conductivity tensor which bears formal resemblance
to the analogous expression (Eq.~\eqref{Eq:frequency-time}) in the case with electron$-$phonon interaction
\begin{equation}
 \label{Eq:frequency-time-imp}
 \sigma_{ab}(t,\omega)=\im\mrd\frac{nq^2}{m\omega}\mrd\delta_{ab}-f_{ab}(t,\omega)-f_{ab}(t,-\omega)^*
\end{equation}
where $f_{ab}(t,\omega)$ is defined as
\begin{equation}
\label{Eq:definition_f-imp}
\begin{split}
 &f_{ab}(t,\omega)=\frac{q^2}{\hbar^2\omega V}\sum_{\alpha\beta\gamma}\e^{\frac{\im}{\hbar}(\epsilon_\alpha-\epsilon_\beta)t}
\langle\beta|\hat\rho_\mathrm{e}|\alpha\rangle \times \\
&\times
A_{\alpha\gamma}(x_{a;\gamma}-x_{a;\alpha})A_{\gamma\beta}(x_{b;\beta}-x_{b;\gamma})
\mathcal{D}(\epsilon_\beta-\epsilon_\gamma+\hbar\omega).
\end{split}
\end{equation}
% while average values $\langle\hat c_\alpha^\dagger\hat c_\beta\rangle$ are given as
% \begin{equation}
%  \langle\hat c_\alpha^\dagger\hat c_\beta\rangle=\delta_{\alpha\beta}\mrd r_\alpha,
% \end{equation}
% where $r_\alpha=\langle\alpha|\hat\rho(0)|\alpha\rangle$ is the occupation of electronic state $\alpha$.
Again, when the initial density matrix $\hat\rho(0)$ is an analytic function of the electronic part of the Hamiltonian $\hat H_{0}$,
the quantity $f_{ab}(t,\omega)$ defined in Eq.~\eqref{Eq:definition_f-imp} contains only populations of the individual electronic
states $r_\alpha=\langle\alpha|\hat\rho(0)|\alpha\rangle$ and does not depend explicitly on time. Thus, $\sigma_{ab}(t,\omega)$
is the optical conductivity tensor (entirely expressed in terms of populations of electronic states).
The final expression for the real part of AC mobility in the presence of interaction with impurities reads
\begin{equation}
\label{Eq:re-mu-imp}
\begin{split}
\mathrm{Re}\mrd\mu_{ab}&(\omega)=\frac{q}{2\hbar\omega}\sum_{\alpha\beta}
(x_{a;\beta}-x_{a;\alpha})(x_{b;\beta}-x_{b;\alpha})
\frac{r_\beta}{\sum_\gamma r_\gamma} \times \\ \times
&\left[w_{\beta\alpha,\mathrm{imp}}(\epsilon_\beta-\epsilon_\alpha+\hbar\omega)- 
      w_{\beta\alpha,\mathrm{imp}}(\epsilon_\beta-\epsilon_\alpha-\hbar\omega)\right],
\end{split}      
\end{equation}
where the terms $w_{\beta\alpha,\mathrm{imp}}$ are of the form
\begin{equation}
 w_{\beta\alpha,\mathrm{imp}}(\epsilon_\beta-\epsilon_\alpha)
=\frac{2\pi}{\hbar}|A_{\alpha\beta}|^2\delta(\epsilon_\beta-\epsilon_\alpha).
\end{equation}
We emphasize the formal analogy between equations~\eqref{Eq:re-mu-eph} and~\eqref{Eq:re-mu-imp}
for the AC mobility. The form of both equations is the same, regardless of the particular interaction mechanism
(electron$-$phonon interaction or interaction with an additional static potential) which causes transitions between localized states.}

\section{Numerical Results}\label{Sec:num}

\subsection{One-dimensional model with Miller-Abrahams rates and Gaussian density of states}\label{Sec:num1d}

In this section, we apply the derived formulas to an one$-$dimensional Gaussian disorder model. The assumption of the model is that the states are located on the sites of an one{$-$}dimensional lattice with spacing $a$ and that the energies of the states are drawn from a Gaussian distribution with standard deviation $\sigma$. The transition rates were assumed to take the Miller-Abrahams form and only the hops between nearest neighbors were considered. Under these assumptions, the transition rate from the state $\beta$ to the state $\gamma$ has the form
\begin{equation}\label{eq:ma_rates}
w_{\beta\gamma}=w_0\mrd\e^{-a/a_{\mathrm{loc}}}
\exp\left(\frac{\epsilon_\beta-\epsilon_\gamma-|\epsilon_\beta-\epsilon_\gamma|}{2k_BT}\right),
\end{equation}
where $a_\mathrm{loc}$ is the localization length which is assumed equal for all sites, $T$ is the temperature, and $w_0$ is a constant prefactor. Real part of the frequency dependent mobility can under all these assumptions be written in the form
\begin{equation}
\label{izrazzaremu}
\mathrm{Re}\mrd\mu_{xx}(\omega)=\sum_{\gamma}\mu_{\gamma,\gamma+1}(\omega),
\end{equation}
where $\mu_{\gamma,\gamma+1}(\omega)$ is the contribution of the pair of sites $(\gamma,\gamma+1)$ given as
\begin{equation}
\label{doprinos}
\mu_{\gamma,\gamma+1}(\omega)=\frac{qa^2}{2k_BT}\mrd w_0\mrd\e^{-a/a_\mathrm{loc}}\mrd M(x).
\end{equation}
In the last equation, $x$ is a dimensionless parameter defined as $x=\beta\hbar\omega$ ($\beta=1/(k_BT)$), while $M(x)$ is the function that reads
\begin{equation}
\label{definicijam}
M(x)=
% \begin{cases}
% \displaystyle{\frac{r_\mathrm{min}}{\sum_\delta r_\delta}\mrd\e^{-x_{\gamma,\gamma+1}}
% \cdot 2\frac{\sinh x}{x}},&x<x_{\gamma,\gamma+1}\\
% \displaystyle{\frac{r_\mathrm{max}}{\sum_\delta r_\delta}
% \frac{1}{x}\left(1-\e^{x_{\gamma,\gamma+1}}\e^{-x}\right)+ \\ +
% \frac{r_\mathrm{min}}{\sum_\delta r_\delta}
% \frac{1}{x}\left(1-\e^{-x_{\gamma,\gamma+1}}\e^{-x}\right)}\quad,
% &x>x_{\gamma,\gamma+1}
% \end{cases}
\begin{cases}
\displaystyle{\frac{r_\mathrm{min}}{\sum_\delta r_\delta}\mrd\e^{-x_{\gamma,\gamma+1}}
\cdot 2\frac{\sinh x}{x}},&x<x_{\gamma,\gamma+1}\\
\begin{split}
\frac{r_\mathrm{max}}{\sum_\delta r_\delta}
\frac{1}{x}\left(1-\e^{x_{\gamma,\gamma+1}}\e^{-x}\right)+ \\ +
\frac{r_\mathrm{min}}{\sum_\delta r_\delta}
\frac{1}{x}\left(1-\e^{-x_{\gamma,\gamma+1}}\e^{-x}\right),
\end{split}
&x>x_{\gamma,\gamma+1}
\end{cases}
\end{equation}
where $x_{\gamma,\gamma+1}=\beta|\epsilon_\gamma-\epsilon_{\gamma+1}|$ and $r_\mathrm{max}$ ($r_\mathrm{min}$) is the population of the state with larger (smaller) energy among the states $\gamma$ and  $\gamma+1$.

The frequency range in which this formula can be applied is determined by the condition \eqref{relax}
where $\tau$ is the relaxation time towards equilibrium. The relaxation time $\tau$ must be larger than the reciprocal value of largest hopping rates $\tau\gtrsim w_0^{-1}\mrd\e^{a/a_\mathrm{loc}}$, so that the relevant frequencies obey the condition
$f\gtrsim (2\pi)^{-1}\mrd w_0\mrd\e^{-a/a_\mathrm{loc}}$.

\begin{figure}[htbp]
\begin{center}
 \includegraphics[width=0.4\textwidth]{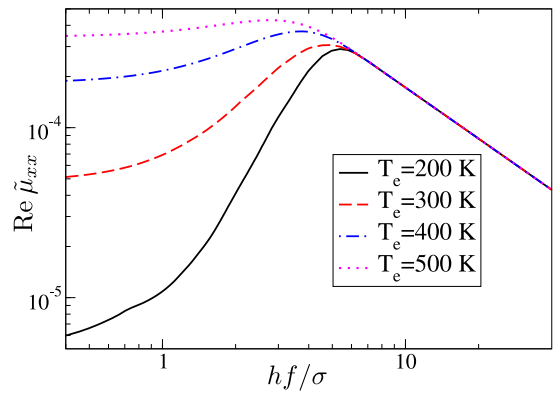}
\end{center}
\caption{(Color online) Frequency dependence of real part of the normalized mobility $\tilde{\mu}_{xx}=\mu_{xx}/\left(  qa^2\beta w_0\e^{-a/a_{loc}}/2\right)$ for different electronic temperatures $T_\mathrm{e}$ in one-dimensional Gaussian disorder model. {The non-equilibrium populations of electronic states were assumed as $r_\gamma=\e^{-\beta_\mathrm{e}\epsilon_\gamma}$.}}
\label{fig:case1}
\end{figure}

\begin{figure}[htbp]
\begin{center}
 \includegraphics[width=0.4\textwidth]{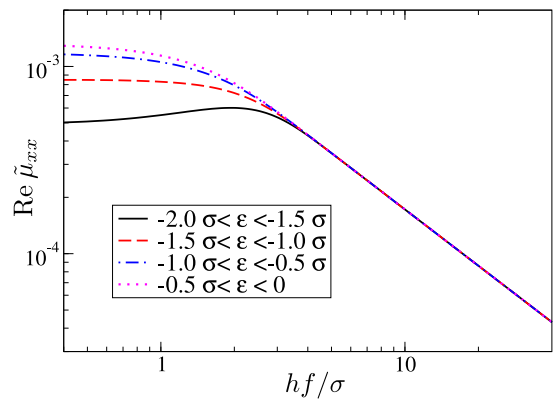}
\end{center}
\caption{(Color online) Frequency dependence of real part of the normalized mobility $\tilde{\mu}_{xx}=\mu_{xx}/\left( qa^2\beta w_0\e^{-a/a_{loc}}/2\right)$ for different choices of the energy interval $\epsilon\in(\epsilon_\mathrm{min},\epsilon_\mathrm{max})$ of the populated states in one-dimensional Gaussian disorder model. {The non-equilibrium populations of electronic states were assumed as $r_\gamma=1$ for $\epsilon_\mathrm{min}<\epsilon_\gamma<\epsilon_\mathrm{max}$, $r_\gamma=0$ otherwise.}}
\label{fig:case2}
\end{figure}

The calculations were performed for a lattice with $10^5$ sites, where the following values of the parameters were used: $T=300\mrd\mathrm{K}$, $\sigma=100\mrd\mathrm{meV}$, $a=1\mrd\mathrm{nm}$, $a_\mathrm{loc}=2a/9$,
$w_0=1.0\times 10^{14}\mrd\mathrm{s}^{-1}$. Two different cases for initial populations of localized states were considered. In case 1, we assume that initial distribution of carriers {are non-equilibrium, but} still of Maxwell-Boltzmann form with an electronic temperatures $T_\mathrm e$ {which can be different than $T$}. Therefore, in this case $r_\gamma=\e^{-\beta_\mathrm{e}\epsilon_\gamma}$, where $\beta_\mathrm{e}=1/\left(k_BT_\mathrm{e}\right)$. In case 2, we assume that only the states in some narrow energy window are initially populated, while the other states are not populated. The initial populations are then given as $r_\gamma=1$ for $\epsilon_\mathrm{min}<\epsilon_\gamma<\epsilon_\mathrm{max}$, $r_\gamma=0$ otherwise. The results for different values of the parameter $T_\mathrm e$ in case 1 and different intervals $(\epsilon_\mathrm{min},\epsilon_\mathrm{max})$ in case 2 are shown in Figs.~\ref{fig:case1} and~\ref{fig:case2}.

As can be immediately seen from expressions in Eqs.~\eqref{izrazzaremu},~\eqref{doprinos} and~\eqref{definicijam}, for sufficiently high frequencies $f$, such that $\displaystyle{hf>\max_{\gamma}
|\epsilon_\gamma-\epsilon_{\gamma+1}|}$, real part of the AC mobility decreases as
$\mathrm{Re}\mrd\mu_{xx}(f)\sim 1/f$. 
On the other hand, for sufficiently low frequencies $f$, such that $\displaystyle{hf<\min_{\gamma}
|\epsilon_\gamma-\epsilon_{\gamma+1}|}$, real part of the mobility tends to a constant value which
depends on the particular choice of $r_\gamma$.

In case 1 and in the intermediate frequency range real part of the AC mobility reaches its maximum value. The height of this maximum (measured relative to the low$-$frequency limit of the mobility) decreases with increasing the temperature $T_\mathrm{e}$. The position of the maximum moves towards lower frequencies with increasing the temperature $T_\mathrm{e}$. Namely, the position of the maximum of the mobility spectrum is determined by the positions of the maximum of the function $M(x)$. For all values of the parameter $T_\mathrm{e}$ considered in Fig.~\ref{fig:case1}, it can be shown (by direct inspection) that the function $M(x)$ has its maximum at $x=x_{\gamma,\gamma+1}.$
At low temperatures $T_\mathrm{e}$ the lowest energy states have the
highest values of the factors $r_\gamma$ and the typical energy difference $|\epsilon_\gamma-\epsilon_{\gamma+1}|$
of the pair of neighboring sites giving significant contribution to the mobility
(at least one of the states should have high enough population factor)
is fairly high, so that the peak of the contribution $\mu_{\gamma,\gamma+1}$ is at high frequencies.
This typical energy difference
%$|\epsilon_\gamma-\epsilon_{\gamma+1}|$ %for the pairs of sites giving significant contribution to the total mobility decreases with increasing
decreases with increasing
the temperature $T_\mathrm e$ (since higher energy states, which are more numerous, also have appreciable values of population factors),
which leads to the shift of the peak position towards lower frequencies. For small enough $|\epsilon_\gamma-\epsilon_{\gamma+1}|$ (compared to
$k_BT$), the function $M(x)$,
for $x<x_{\gamma,\gamma+1}$,
can be approximated by a constant, which leads to flattening of the maximum, as seen at higher electronic temperatures in Fig.~\ref{fig:case1}.

A similar analysis can be used to understand the shapes of the mobility spectra for case 2 shown in Fig.~\ref{fig:case2}. The contribution to the mobility of the pair $(\gamma,\gamma+1)$ reaches its maximum at frequency $f_*$ such that $hf_*>|\epsilon_{\gamma}-\epsilon_{\gamma+1}|$. When the interval $(\epsilon_\mathrm{min},\epsilon_\mathrm{max})$ is in the tail of the Gaussian, the typical energy difference $|\epsilon_\gamma-\epsilon_{\gamma+1}|$ is rather high for the pairs contributing significantly to the mobility, so that the maximum of the mobility spectrum is at high frequencies. Moving the interval towards the center of the Gaussian, the typical energy difference decreases and so does the position of the maximum of the mobility spectrum. For sufficiently small energy difference (compared to $k_BT$), the function $M(x)$ can be well approximated by a constant in the range $x<x_{\gamma,\gamma+1}$ which explains the disappearance of the maximum.

{Since the flattening of the maximum in the mobility spectrum appears due to presence of carriers at higher energies under non-equilibrium conditions, this flattening may be considered as a signature of non-equilibrium effects in the system. It is less pronounced when non-equilibrium distribution is of Maxwell-Boltzmann type with a different electronic temperature and more pronounced in the case when the carriers are present only at energies in a certain spectral window - a situation where the carrier distribution more strongly differs from the equilibrium one.}

\subsection{Model of a disordered conjugated polymer material}

Next, we apply the derived formula for frequency dependence of the mobility to a realistic polymer material -- strongly disordered poly(3-hexylthiophene) (P3HT) polymer. The positions of electronic states, hopping probabilities between the states and the energies of states were extracted from our previous calculations reported in Ref.~\onlinecite{jpcb115-1792}. For completeness, we briefly summarize the methodology employed in these calculations. First, the positions of atoms were obtained from classical molecular dynamics simulations using a simulated annealing procedure. 50 different realizations of the 5~nm$\times$5~nm$\times$5~nm portion of material (that consists of 12024 atoms) were obtained from these simulations and were subsequently used in electronic structure calculations. Charge patching method\cite{jcp128-121102} was used to obtain the single-particle Hamiltonian that approximates well the Hamiltonian that would be obtained from density functional theory in local density approximation. This 
Hamiltonian was diagonalized using the overlapping fragments method.\cite{jcp134-094119} 
The transition rates for downhill transitions between the states were then calculated as
\begin{equation}\label{eq:yuk-aa}
w_{\alpha\beta}=\alpha^2 \mathcal{S}_{\alpha\beta}^2 \left[N(\epsilon_{\alpha\beta})+1\right]D_{ph}(\epsilon_{\alpha\beta})/\epsilon_{\alpha\beta},
\end{equation}
where $D_{ph}(E)$ is the phonon DOS normalized such that $\int_0^\infty D_{ph}(E)\mathrm{d}E=1$, $\epsilon_{\alpha\beta}=|\epsilon_\alpha-\epsilon_\beta|$,  $N(E)$ is the phonon occupation number given by the Bose-Einstein distribution at a temperature $T$, $\mathcal{S}_{\alpha\beta}=\int \mathrm{d}^3{\mathbf{r}}|\psi_\alpha({\mathbf{ r}})|\cdot|\psi_\beta({\mathbf {r}})|$ is the overlap of the wavefunction moduli and $\alpha$ is a constant factor equal to $10^7\:\textrm{eV}\textrm{s}^{-1/2}$. {The phonon energies and the phonon DOS were calculated from the classical force field that was used in molecular dynamics simulations by diagonalizing the corresponding dynamical matrix, as reported in Ref.~\onlinecite{nl9-3996}.}  Eq.~\eqref{eq:yuk-aa} gives a good approximation of the transition rates that would be obtained from Eq.~\eqref{femijevozp}, as shown in Ref.~\onlinecite{apl97-043305}. The value of the parameter $\alpha$ in Eq.~\eqref{eq:yuk-aa} was chosen to provide the best fit of Eq.~\eqref{eq:yuk-aa} 
to Eq.~\eqref{femijevozp}.

\begin{figure}[htbp]
\begin{center}
 \includegraphics[width=0.4\textwidth]{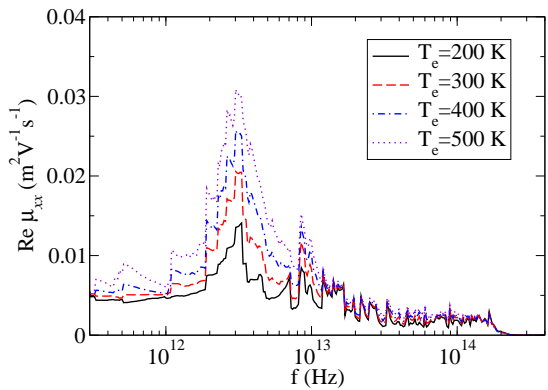}
\end{center}
\caption{(Color online) Frequency dependence of real part of hole mobility in disordered P3HT polymer for different electronic temperatures $T_\mathrm{e}$ and the lattice temperature $T=300\:\mathrm{K}$.}
\label{fig:mup3ht}
\end{figure}

Frequency dependence of real part of hole mobility was then calculated using Eqs.~\eqref{opstislucajfzpneravnotezna} and \eqref{eq:yuk-aa} where all data from 50 different realizations of the 12024 atom system were used. The results obtained from the calculation are presented in Fig.~\ref{fig:mup3ht}. We note that the data from electronic structure calculations that were performed are not sufficient to yield convergent results for the mobility. This can be evidenced from the noisy dependence in Fig.~\ref{fig:mup3ht} and from the fact that the mobility obtained from smaller number of realizations of the system is different than the one in Fig.~\ref{fig:mup3ht}. Larger number of calculations or the calculations performed on larger systems would be needed to obtain converged value of the mobility. However, such calculations require a huge computational cost and we cannot currently perform them. Nevertheless, from the set of calculations that were performed one can identify the main trends in the frequency 
dependence of the mobility. As in the simple model discussed in Sec.~\ref{Sec:num1d}, real part of the mobility exhibits a peak at a frequency that corresponds to typical transition energies in the system, which is then followed by a decay at higher frequencies. The mobility also increases with an increase in electronic temperature, as expected.

{We next compare the results obtained in this work to measurements of high-frequency P3HT hole mobility reported in the literature. These measurements are typically based on time-resolved terahertz spectroscopy\cite{rmp83-543} and cover the frequencies around 1 THz. At these frequencies our simulations yield the mobilities on the order of $(50-100)\:\mathrm{cm}^2/(\mathrm{V}\cdot\mathrm{s})$. In Ref.~\onlinecite{jpcb110-25462} the mobilities on the order of $10\:\mathrm{cm}^2/(\mathrm{V}\cdot\mathrm{s})$ were extracted from the fits to measurements. On the other hand, the mobilities on the order of $50\:\mathrm{cm}^2/(\mathrm{V}\cdot\mathrm{s})$ were obtained in Ref.~\onlinecite{jpcc112-7928}. Therefore, the simulation yields the same order of magnitude of the terahertz mobility as previously reported in experiments.}

\section{Discussion}\label{Sec:disc}
                          
In this Section, we discuss our results in light of other results that exist in the literature and concern optical conductivity in system with localized states.

Our result for non-equilibrium optical conductivity should, of course, in the special case of equilibrium reduce to the formula valid in equilibrium case. A well-known expression for real part of optical conductivity  in equilibrium that relates it to the mean square displacement of a diffusing carrier reads (see for example Ref.~\onlinecite{prb83-081202})
\begin{equation}\label{eq:kuboquantum}
\mathrm{Re}\:\sigma\left(\omega\right)=-\frac{q^2\omega^2}{V}\frac{\tanh{\left(\beta\hbar\omega/2\right)}}{\hbar\omega}\mrd\mathrm{Re}\int_0^{+\infty} \mathrm{d}t\mrd \e^{\im\omega t}\Delta X^2(t),
\end{equation}
where $\Delta X^2(t)=\left\langle \left(\hat{X}(t)-\hat{X}(0)\right)^2 \right\rangle$, $\hat{X}$ is the sum of position operators of all electrons
$\langle...\rangle=\mathrm{Tr}\left(\e^{-\beta\hat H}...\right)/\mathrm{Tr}\mrd\e^{-\beta\hat H}$ is the thermodynamic average
at the temperature $T=1/(\beta k_B)$. 
While, at first sight, Eq.~\eqref{eq:kuboquantum} 
seems to lead to rather different result for the lowest$-$order optical conductivity than the one embodied in Eq.~\eqref{opstislucajfzpneravnotezna},
a detailed proof can be conducted, showing that
the two expressions
are identical for the system with localized states in equilibrium. The details of this proof are given in Appendix \ref{Sec:App}.

A somewhat different version of Eq.~\eqref{eq:kuboquantum} is often encountered in the literature which contains the $\beta/2$ term instead of the $\frac{\tanh{\left(\beta\hbar\omega/2\right)}}{\hbar\omega}$ term and reads\cite{prb7-4491,jppa215-123,jpcc116-19665,irpc27-87} 
\begin{equation}\label{eq:kuboclassical}
\mathrm{Re}\:\sigma\left(\omega\right)=-\frac{q^2\omega^2\beta}{2V}\mathrm{Re}\int_0^{+\infty} \mathrm{d}t\mrd \e^{\im\omega t}\Delta X^2(t).
\end{equation}
When the condition $\beta\hbar\omega\ll 1$ is satisfied these two expressions are approximately equal. However, at high frequencies these two expressions essentially differ. While Eq.~\eqref{eq:kuboquantum} leads to real part of the conductivity that vanishes at sufficiently high frequencies, Eq.~\eqref{eq:kuboclassical} gives a constant real part of the mobility at these frequencies which is not the correct trend. Therefore, Eq.~\eqref{eq:kuboclassical} should be applied only if the condition $\beta\hbar\omega\ll 1$ is satisfied.

An expression for optical conductivity in the form similar to the one given in Eq.~\eqref{opstislucajfzpneravnotezna} has also been previously obtained for the case of equilibrium.\cite{pssb65-665,jpcm1-557} These expressions [Eq. (12) in Ref.~\onlinecite{pssb65-665} and Eq. (3.21) in Ref. \onlinecite{jpcm1-557}] in the limit of low concentration are the special case of Eq.~\eqref{opstislucajfzpneravnotezna} for the case of equilibrium in the limit $\hbar\omega\ll k_BT$. It is very interesting that our main result given by Eq.~\eqref{opstislucajfzpneravnotezna} has the same mathematical form as 
the expressions for the case of equilibrium. Therefore, we have generalized the result that was known for the case of equilibrium to the case of non-equilibrium systems that satisfy the assumptions of factorization of the density matrix into the electron and the phonon part [Eq. \eqref{statfakt}] and weak relaxation at relevant timescales [Eq. \eqref{relax}].

As we have already pointed out, our results are not expected to be valid at low frequencies, such that the period of perturbation is larger than the carrier relaxation time. For that reason, one can certainly not assume that DC mobility or conductivity is equal to the low frequency limit of our results. There is an additional reason that our results cannot be extended to low frequencies. It has been pointed out in Refs.~\onlinecite{pssb65-665,jpcm1-557} that conductivity at low frequencies cannot be obtained from a formal expansion in powers of electron-phonon interaction strength, which is an approach used in our work.

From the previous discussion, we can conclude that our results reduce to previous results from the literature for the case of equilibrium state. On the other hand, there have been almost no works in the literature with an attempt to obtain similar results for the system out of equilibrium. The exceptions are Refs. \onlinecite{prl93-250601,arxiv0404270} where Eq.~\eqref{kubojj} was derived. However, we are not aware of any attempt to obtain a more specific form of non-equilibrium conductivity in a system with localized states and the main contribution of our work is that it covers this so far unexplored area.

\section{Conclusion}    \label{Sec:Conclusion}                               
In conclusion, we have developed an approach for the treatment of non-equilibrium optical conductivity in a system with localized electronic states and weak electron-phonon or electron-impurity interaction. Starting from non-equilibrium generalization of Kubo's formula and performing the expansion of optical conductivity in powers of small electron-phonon interaction parameter, we obtain a relatively simple expression for the optical conductivity of the material. {In the special case of incoherent non-equilibrium state} the expression contains only the positions of electronic states, their non-equilibrium populations and Fermi's golden rule transition probabilities between the states. Interestingly, the same mathematical form of the expression is valid both in the case of electron-phonon and electron-impurity interaction. 
Our result opens the way to better understanding of the response of non-equilibrium systems to electromagnetic radiation. A typical example where our results can be applied is photoexcited semiconductor where electrons and holes are formed by the optical excitation. If that semiconductor is then probed by low energy (terahertz) excitation, the response will depend on the non-equilibrium distribution of excited carriers. Our final expressions should be able to predict the response of the system to such probes. The application of the derived formula to two model systems was presented to illustrate the features that one may expect to see in terahertz conductivity spectra.

\section{Acknowledgments}
This work was supported by European Community FP7 Marie Curie Career Integration Grant (ELECTROMAT), Serbian Ministry of Education, Science and Technological Development (project ON171017) and FP7 projects PRACE-3IP and EGI-InSPIRE.
V. J. also acknowledges the support by the Fund for Young Talents of the Serbian Ministry of Youth and Sport.

\appendix 
\section{Proof of equivalence of the lowest$-$order optical conductivity calculated from
Eq.~\eqref{eq:kuboquantum} and the expression in Eq.~\eqref{opstislucajfzpneravnotezna}}\label{Sec:App}

The operator $\hat X(t)-\hat X(0)$ appearing in Eq.~\eqref{eq:kuboquantum} can be expressed as (see Eqs.~\eqref{dipolni} and~\eqref{jednkont})
\begin{equation}\label{eq:positiondif}
 q\left(\hat X(t)-\hat X(0)\right)=\int_0^t\dif t'\mrd\hat{J}_x(t'),
\end{equation}
so that in the case of localized carriers, when $\hat J_x=\hat J_x^{(2)}$, the operator $\left(\hat X(t)-\hat X(0)\right)^2$
is quadratic in electron$-$phonon coupling constants. If we are to obtain the conductivity up to quadratic terms in small interaction
constants $g_{\alpha\beta,k}^\pm$, it is clear that the following factorization of the equilibrium statistical operator should be adopted
(compare to the decomposition of the initial density matrix in Eq.~\eqref{statfakt})
\begin{equation}
\label{Eq:dekompozicija}
 \frac{\e^{-\hat H/(k_BT)}}
{\mathrm{Tr}\mrd\e^{-\hat H/(k_BT)}}\approx\frac{\e^{-\hat H_\mathrm e/(k_BT)}}
{\mathrm{Tr}_\mathrm e\mrd\e^{-\hat H_\mathrm e/(k_BT)}}
\frac{\e^{-\hat H_\mathrm{ph}/(k_BT)}}
{\mathrm{Tr}_\mathrm{ph}\mrd\e^{-\hat H_\mathrm{ph}/(k_BT)}},
\end{equation}
and that time dependencies appearing in~\eqref{eq:kuboquantum} should be taken with respect to the non$-$interacting Hamiltonian $\hat H_0$.
The average value $\left\langle \left(\hat{X}(t)-\hat{X}(0)\right)^2 \right\rangle$ is then transformed into
\begin{equation}\label{Eq:disperzija1c}
\begin{split}
 &\left\langle\left(\hat{X}(t)-\hat{X}(0)\right)^2\right\rangle=\\
&\sum_{k}\sum_{\alpha\beta\gamma\delta}(x_\beta-x_\alpha)(x_\delta-x_\gamma)
\langle\hat c_\alpha^\dagger\hat c_\beta\hat c_\gamma^\dagger\hat c_\delta\rangle_\mathrm{e}\times\\
&\left(
g_{\alpha\beta,k}^-g_{\gamma\delta,k}^+
\frac{\e^{\frac{\im}{\hbar}(\epsilon_\alpha-\epsilon_\beta-\hbar\omega_k)t}-1}{\epsilon_\alpha-\epsilon_\beta-\hbar\omega_k}
\frac{\e^{\frac{\im}{\hbar}(\epsilon_\gamma-\epsilon_\delta+\hbar\omega_k)t}-1}{\epsilon_\gamma-\epsilon_\delta+\hbar\omega_k}
\mrd(1+N_k)+ \right. \\  &\left.+
g_{\alpha\beta,k}^+g_{\gamma\delta,k}^-
\frac{\e^{\frac{\im}{\hbar}(\epsilon_\alpha-\epsilon_\beta+\hbar\omega_k)t}-1}{\epsilon_\alpha-\epsilon_\beta+\hbar\omega_k}
\frac{\e^{\frac{\im}{\hbar}(\epsilon_\gamma-\epsilon_\delta-\hbar\omega_k)t}-1}{\epsilon_\gamma-\epsilon_\delta-\hbar\omega_k}
\mrd N_k\right),
\end{split}
\end{equation}
where $\langle...\rangle_\mathrm{e}$ denotes averaging with respect to the electronic part of the decomposition~\eqref{Eq:dekompozicija}.
Combining equations~\eqref{eq:kuboquantum} and~\eqref{Eq:disperzija1c} and in the limit of low carrier densities,
when Eqs.~\eqref{4u1},~\eqref{analiticka} and~\eqref{ralpha} can be used,
the following expression for the optical conductivity ($\omega\neq 0$) is obtained
\begin{equation}\label{Eq:ravnotezafzp}
 \begin{split}
  &\mathrm{Re}\mrd\sigma_{xx}(\omega)=
\frac{q^2}{2\hbar\omega V}\tanh\left(\frac{\hbar\omega}{2k_BT}\right)\sum_{\alpha\beta}(x_\beta-x_\alpha)^2 r_\beta\times\\
&\times[w_{\beta\alpha,\mathrm{ph}}(\epsilon_\beta-\epsilon_\alpha-\hbar\omega)+
w_{\beta\alpha,\mathrm{ph}}(\epsilon_\beta-\epsilon_\alpha+\hbar\omega)]
 \end{split}
\end{equation}
where the transition probabilities $w_{\beta\alpha,\mathrm{ph}}$ are defined in Eq.~\eqref{femijevozp} and
the average occupation of electronic state $\beta$ is
$r_\beta=\e^{-\epsilon_\beta/(k_BT)}/\mathrm{Tr}_\mathrm{e}\mrd\e^{-\hat H_\mathrm{e}/(k_BT)}$.
In order to prove that
the relation~\eqref{opstislucajfzpneravnotezna} (in which we take
$r_\beta=\e^{-\epsilon_\beta/(k_BT_\mathrm{ph})}/\mathrm{Tr}_\mathrm{e}\mrd\e^{-\hat H_\mathrm{e}/(k_BT_\mathrm{ph})}$) gives the
same result for the lowest$-$order optical conductivity as the equation~\eqref{Eq:ravnotezafzp} (assuming that $T=T_\mathrm{ph}$), we note that the transition probabilities satisfy
the detailed$-$balance condition (in the low$-$density limit and in the presence of external harmonic perturbation)
\begin{equation}\label{Eq:detaljnaperturbacija}
\begin{split}
 \frac{w_{\beta\alpha,\mathrm{ph}}(\epsilon_\beta-\epsilon_\alpha+\hbar\omega)}
{w_{\alpha\beta,\mathrm{ph}}(\epsilon_\alpha-\epsilon_\beta-\hbar\omega)}&=
\e^{-(\epsilon_\alpha-\epsilon_\beta-\hbar\omega)/(k_BT)}=\\
&=\frac{r_\alpha}{r_\beta}\frac{1+\tanh\frac{\hbar\omega}{2k_BT}}{1-\tanh\frac{\hbar\omega}{2k_BT}}.
\end{split}
\end{equation}
Interchanging the dummy electronic indices $\alpha,\beta$ in the first summand in the Eq.~\eqref{opstislucajfzpneravnotezna} we obtain
\begin{equation}
\label{Eq:maopstifzp}
 \begin{split}
  &\mathrm{Re}\mrd\sigma_{xx}(\omega)=
\frac{q^2}{2\hbar\omega V}\sum_{\alpha\beta}
(x_\beta-x_\alpha)^2\times\\
&\times[-r_\alpha w_{\alpha\beta,\mathrm{ph}}(\epsilon_\alpha-\epsilon_\beta-\hbar\omega)
+r_\beta w_{\beta\alpha,\mathrm{ph}}(\epsilon_\beta-\epsilon_\alpha+\hbar\omega)]
 \end{split}
\end{equation}
whereas performing the same operation on the Eq.~\eqref{Eq:ravnotezafzp} gives
\begin{equation}
 \label{Eq:maravnotezafzp}
 \begin{split}
  &\mathrm{Re}\mrd\sigma_{xx}(\omega)=
\frac{q^2}{2\hbar\omega V}\tanh\left(\frac{\hbar\omega}{2k_BT}\right)\sum_{\alpha\beta}(x_\beta-x_\alpha)^2\times\\
&\times[r_\alpha w_{\alpha\beta,\mathrm{ph}}(\epsilon_\alpha-\epsilon_\beta-\hbar\omega)
+r_\beta w_{\beta\alpha,\mathrm{ph}}(\epsilon_\beta-\epsilon_\alpha+\hbar\omega)].
 \end{split}
\end{equation}
The right$-$hand sides of the Eqs.~\eqref{Eq:maopstifzp} and~\eqref{Eq:maravnotezafzp} are equal since, by the condition~\eqref{Eq:detaljnaperturbacija},
single summands under the double sums are mutually equal.

% \newpage 
%

\end{document}